\documentclass[twocolumn,tighten,twocolappendix]{aastex631}

\usepackage{graphicx}
\usepackage{url}
\usepackage{hyperref}
\usepackage{amsmath}
\usepackage{amssymb}
\usepackage{xspace}

\DeclareGraphicsExtensions{.pdf,.png,.jpeg}

\definecolor{ForestGreen}{rgb}{0.133,0.545,0.133}

\def\rp{r_\gamma}
\def\m87{M87*}
\def\sgra{\mbox{Sgr A*}\xspace}

\begin{document}

\title{Measuring the Shape of Kerr Black Holes at the Photon Orbit}

\author[0009-0006-0070-1888]{Kiana Salehi}
\affiliation{Perimeter Institute for Theoretical Physics, 31 Caroline Street North, Waterloo, ON, N2L 2Y5, Canada}
\affiliation{Department of Physics and Astronomy, University of Waterloo, 200 University Avenue West, Waterloo, ON, N2L 3G1, Canada}
\affiliation{Waterloo Centre for Astrophysics, University of Waterloo, Waterloo, ON N2L 3G1 Canada}

\author[0000-0002-3351-760X]{Avery E. Broderick}
\affiliation{Perimeter Institute for Theoretical Physics, 31 Caroline Street North, Waterloo, ON, N2L 2Y5, Canada}
\affiliation{Department of Physics and Astronomy, University of Waterloo, 200 University Avenue West, Waterloo, ON, N2L 3G1, Canada}
\affiliation{Waterloo Centre for Astrophysics, University of Waterloo, Waterloo, ON N2L 3G1 Canada}

\begin{abstract}
The bright ring-like structures observed in the images of \m87 and \sgra captured by the Event Horizon Telescope strongly support the validity of general relativity.
Lensed images of the emission region, often referred to as photon rings in this context, are a direct consequence of the unstable dynamics of null geodesics near the spherical photon orbit in the Kerr spacetime.
The order of the lensed image can be characterized by the number of half-orbits the photons complete before reaching the observer, with higher-order photon rings produced by null geodesics that circle the black hole more times. 
However, low-order rings are significantly influenced by the astrophysical environment.
Measuring the Lyapunov exponent requires probing the exponentially small differences between successive photon rings or between photon rings and the shadow. 
We investigate 
potential astrophysical sources of systematic error the estimation of Lyapunov exponent, including the location of the observed emission, and especially at low photon ring order.
We show that it is nevertheless
possible to measure this purely gravitational quantity to roughly 10\% and 1\% systematic uncertainty by resolving the $n=2$ and $n=3$ photon rings with the shadow size, respectively.
Therefore, the forthcoming black hole imaging efforts to capture, even if indirectly, the $n=2$ photon ring can result in a measurement of the Lyapunov exponent that is not limited by astrophysical uncertainties.
\end{abstract}


\section{Introduction} \label{sec:intro}

A new  era in the study of black holes near their horizons has been opened by recent observations.
The Event Horizon Telescope (EHT) has provided the first-ever image of a black hole's event horizon, capturing the shadow of the supermassive black holes in the galaxy \m87 and our own galaxy, 
\sgra \citep{M87_PaperI,M87_PaperII,M87_PaperIII,M87_PaperIV,M87_PaperV,M87_PaperVI,SgpaperI,SgpaperII,SGpaperIII,SgpaperIV,SgpaperV,SgpaperVI}.
In addition, the Laser Interferometer Gravitational-Wave Observatory (LIGO) has enabled the direct detection of gravitational waves. This offers valuable insights into black hole mergers and neutron star collisions \citep{LIGOQNM}.
Moreover, the next-generation Event Horizon Telescope (ngEHT) aims for more detailed observations of black hole dynamics by enhancing the resolution and sensitivity in their observations \citep{Doeleman2019,Johnson2020}.
Together, these observatories are opening up new frontiers in our understanding of gravity in the strong-field regime.

General relativity predicts a central brightness depression for the black holes in Kerr spacetime which is known as the black hole shadow \citep{Hilbert1917}. This is a direct consequence of the existence of an unstable but bound critical photon shell around the Kerr black hole. This shadow is encompassed by an infinite sequence of photon rings, each associated with higher-order images which marginally get closer to the shadow. These rings are results of strong gravitational lensing effect which forces null geodesics to complete additional half-orbits before escaping outwards. The first-order photon orbit corresponds to the null ray that completes one half-orbit, while the second-order orbit completes two half-orbits, and so on \citep{Darwin1959, Luminet1979,BroderickLoeb2006,Johnson2020}. The image of the $n=0$ ring is referred to as the primary image. The $n=1$ ring creates the secondary image, followed by the tertiary image for $n=2$, etc.

Extensive research has been conducted on measuring the shadow size of black holes in the images of \m87 and \sgra which provides valuable insights into the characteristics of Kerr spacetime, such as the value of the spin and mass. However, shadow size measurement is not the only observable that can be helpful.
Improving resolution in the future observations will allow us to resolve subsequent photon rings \citep{Chael2016,Dexter2009, Moscibrodzka2009}. Due to their highly lensed and unstable nature, the relative radii of these higher-order photon rings can be characterized by a Lyapunov exponent. This quantity fundamentally a measure of the tidal instability along the photon orbit and is proportional to the square root of the second derivative of the effective potential at the circular photon orbit, measuring how fast small radial perturbations grow. This quantity represents another promising observable that could further elucidate the characteristics of black holes.  

One of the primary sources of uncertainty in the measurement of the Lyapunov exponent arises from its calculation as a means to characterize the unstable nature of the relative radii of high-order photon rings, which represents a purely gravitational effect. While it is possible to measure the shadow size, the individual photon rings remain spatially unresolved. Although achieving the precision required to resolve high-order photon rings may not be feasible in the near future, the prospect of resolving low-order photon rings within the next few decades remains plausible \citep{Johnson2020,Broderick2022,Broderick2022b}. Consequently, one can estimate the value of the Lyapunov exponent using low-order photon rings, but the accuracy of such estimates is a subject we aim to address in this work. Additionally, this is not the only source of uncertainty in measuring the Lyapunov exponent; there are significant challenges related to the precise definition of each of these rings \citep{M87_PaperVI}.

Another main source of uncertantity in estimating the Lyapunov exponen orginates from the deviation between a purely geometric approximation and observational radius of the photon rings and their appearance within an astrophysical context, which is associated with an emission region.  In the former, we identify that null geodesic that executes $n/2$ complete orbits with the the $n$th photon ring, and the $(n+1)$th image.  That is, the $n=1$ photon ring is attributed to the photon trajectory that orbits exactly $\pi$ radians around the back of the black hole prior to reaching the observer at infinity, the $n=2$ photon ring is associated with the trajectory that executes a complete orbit, and so forth.  This definition was particularly useful when considering the properties of photon rings in a large family of non-Kerr spacetimes \citep{SphericalShadows23,Salehi2023}.

However, the astronomical manifestation of photon rings is inevitably linked with the emission of the photons being observed, presumably within an accretion disk in the immediate vicinity of the black hole \citep{Narayan1995,Shakura1973}.  The $n=1$ photon ring is, then, associated with all photons that contribute to the secondary image, and thus the collection of photon trajectories.  For equatorial emission regions, these trajectories will execute between $1/4$ and $3/4$ of a full orbit, intersecting with the emission region exactly twice (see \autoref{fig:n=1 astrophysical picture}).   Similarly, the $n=2$ photon ring is produced by trajectories that intersect the emission region three times (completing between $3/4$ and $5/4$ complete orbits).  This more practical notion of the photon rings necessarily expands the range of the observed radii of the $n$th photon ring around the geometric notion, and provides a new source of astrophysical systematic uncertainty, beyond observational limitations\citep{Yuan2014}. 

While in \citet{SphericalShadows23} and \citet{Salehi2023}, we presented a non-parametric framework for exploring the implications of observing photon rings and black hole shadows for a large family of spacetimes, here we restrict ourselves to Kerr.  This is motivated by both practical and epistemological considerations.  The computations we present --- a quantitative assessment of the magnitude of astrophysical uncertainties on the measurement of the spacetime geometry near the photon shell --- require an explicit specification of the metric, for which Kerr is the natural choice.  However, because general relativity provides a well-established and extensively tested framework that aligns with all current observational data, the Kerr spacetime also represents the natural null hypothesis.  Therefore, in the absence of some as yet unidentified reason to exclude general relativity, deviations from Kerr inferred from photon ring measurements will only be relevant if they exceed the expected uncertainties as estimated within Kerr.  That is, Kerr most naturally provides both the expectation and the yardstick within which to test it.

For practical reasons we will restrict our attention to polar observers.  This is consistent with the near-polar viewing geometry for \m87.  While the inclination of \sgra remains poorly constrained, it lies behind an interstellar scattering screen that blurs its image; even at $1~{\rm THz}$, well above the highest frequency we might entertain from the ground, diffractive scattering blurs the image between $0.6~\mu{\rm as}$ and $1.2~\mu{\rm as}$, ostensibly eliminating efforts to resolve any but the $n=1$ photon ring.  All other prospective horizon-scale imaging targets are significantly smaller, leaving \m87 as the primary astronomical source for current and future photon ring studies.


The formatting of this paper is as follows. In section 2, we briefly review the general relativity's predictions about the multiple photon rings in Kerr spacetime observed by a polar observer. Then we continue the discussion on section 3, by exploring the affects of the astrophysical surroundings on the first few highly lensed images. Section 4 contains discussions on observational implications and expected error bars. Finally we conclude in section 6.

\section{Photon Rings in the Kerr Spacetime}
\label{sec:polar}

\subsection{Trajectories Near the Photon Orbit}

General relativity predicts that all null bound orbits in Kerr spacetime lie at a specific fixed value of r in the range of 
\begin{equation}
    r^\gamma _- \le r  \le r^\gamma _+ 
\end{equation}
where
\begin{equation}
    r^\gamma _\pm  = 2 M \left[1+ \cos \left( \frac{2}{3}  \arccos \left(\pm \frac{a}{M}\right) \right)\right].
\end{equation}
$M$ and $a$ are the mass and spin of black hole, respectfully \citep{Bardeen1972}. In this work we are considering few simplifying assumptions motivated by the EHT observations of M87*:
\begin{enumerate}
\item The observer is located at infinity along side the polar axis in Kerr spacetime.
\item The black hole is surrounded with an astrophysical disk that is positioned at the equatorial plane.
\item The emission is axisymmetric, i.e., produced by rings, and reaches a maximum at a single radius.
\end{enumerate}
Each of the above may be relaxed in future studies.

Thus, the critical radius for the null bound orbits take the form that had been presented in Equation 17 of \citet{Salehi2023} ,
\begin{equation}
\label{eq:polar radius}
    r_\gamma = \frac{N(r_\gamma)}{N'(r_\gamma)}.
\end{equation}
where for the Kerr spacetime 
\begin{equation}
     N^2(r,a)= \frac{\Delta r^2}{(r^2+a^2)^2}.
\end{equation}
Thus, \autoref{eq:polar radius} reduces to a single value of  
\begin{equation}
    r_\gamma = M + 2 M \sqrt{1- \frac{a^2}{3 M^2} }\cos \left[\frac{1}{3} \arccos \left(\frac{1-a^2/M^2}{(1-a^2/3M)^{3/2}}\right)\right].
\end{equation}
These bound orbits are inherently unstable; even small disturbances can cause them to either fall into the black hole or escape to infinity, where they become visible to telescopes. 

We can examine two geodesics, one of which is bound, while the other initially deviates by a small radial separation denoted as $\delta r_0$. According to the equation of geodesic deviation, after undergoing n half-orbits (see \autoref{fig:geomttric pics of n=0,1,2}), their separation undergoes a growth to,
\begin{equation}
\label{eq:delta rn}
    \delta r_n \approx e^{\gamma n} \delta r_0.
\end{equation}
where $\gamma$ is a Lyapunov exponent which for the special case of a polar observer follows the equation 25 of \citet{Salehi2023} \citep[see also,][]{ Ferrari1984,Cardoso2009},
\begin{equation}
    \gamma \equiv 
    \frac{N^{3/2}(\rp,a)}{N'(\rp,a)}
    \left[-\frac{N''(\rp,a)}{B^2(\rp,a)}\right]^{1/2}
    2 K[(a/R)^2],
    \label{eq:gamma}
\end{equation}
\\
where $K(k)$ is the complete elliptic integral of the first kind \citep[see, e.g.,][]{Abramowitz1972} and for Kerr spacetime
\begin{equation}
   B^2(r,a)= \frac{r^4}{(r^2+a^2)^2},
\end{equation} 
where $\Delta = r^2 + a^2 - 2 M r$ and $N'$ and $N''$ are the first order and second order derivatives of the function $N$ with respect to the radial component. 

These photon rings, after completing $n$ half-orbits along trajectories very close to the shadow radius of the black hole, ultimately reach sufficiently large distances such that they can freely escape to infinity ($z \to \infty$). The apparent radii of these orbits on the observer’s image plane are given by
\begin{equation}
    R_n = R_\infty + \Delta R \, e^{-n \gamma},
    \label{eq:Rn and shadow}
\end{equation}
where $R_\infty$ denotes the asymptotic shadow radius.  

From this expression, one immediately finds
\begin{equation}
\label{eq:gamma with R}
   \lim_{n \to \infty} \frac{R_n - R_\infty}{R_{n+1} - R_\infty} = e^{\gamma},
\end{equation}
where the limit part comes from the fact that \autoref{eq:delta rn} becomes more accurate as we go for higher n values.
Moreover, the need for an independent shadow-size measurement may be circumvented if three photon ring radii are observed, since in that case
\begin{equation}
\label{eq:gamma012}
    \gamma = \lim_{n \to \infty} \ln \left( \frac{R_{n+1} - R_n}{R_{n+2} - R_{n+1}} \right).
\end{equation}


\subsection{The Geometric Picture of Photon Rings}
A critical aspect to emphasize here is the definition of subsequent photon rings. The observed black hole image consists of multiple image orders, each appearing as a bright ring on the image plane with a specific radius. These rings are characterized by the number of half-orbits completed before the photons reach the observer. In other words, they undergo a certain number of $\pi$ rotations during their full trajectory, as illustrated by the dashed lines in \autoref{fig:geomttric pics of n=0,1,2}.  

In this figure, the solid dot marks the emission location, while the dashed lines represent the continuation of the trajectory to infinity, as if the ray originated from a distant source. This continuation provides a useful way to identify the order of each ring.  

As shown, each ring order corresponds to a specific emission location on the equatorial plane at $z=0$, indicated by the solid dots. Consequently, each successive ring has a well-defined image radius as measured by an observer at $z \to \infty$. We refer to this as the \textbf{geometric} photon ring radius, denoted by $R^{\mathrm{geo}}_n$.  

To estimate the Lyapunov exponent, one can use these geometric image radii in either \autoref{eq:gamma012} or \autoref{eq:gamma with R}. We define the corresponding geometric estimates as
\begin{equation}
\label{eq:gamma geo n with R}
  \gamma^{\mathrm{geo},01\infty}_n = \ln \left( \frac{R^{\mathrm{geo}}_n - R_\infty}{R^{\mathrm{geo}}_{n+1} - R_\infty} \right),
\end{equation}
and
\begin{equation}
\label{eq:gamma geo n no R}
  \gamma^{\mathrm{geo},012}_n = \ln \left( \frac{R^{\mathrm{geo}}_n - R^{\mathrm{geo}}_{n+1}}{R^{\mathrm{geo}}_{n+2} - R^{\mathrm{geo}}_{n+1}} \right).
\end{equation}

It is immediately clear that deviations arise between these geometric estimates of the Lyapunov exponent and the expression in \autoref{eq:gamma}, particularly when using low-order rings such as $n=1$ or $2$. However, this is not the only potential source of error in estimating the Lyapunov exponent; another important source will be discussed in the next subsection and explored further in the following section.

\begin{figure*}
    \centering    
    \includegraphics[width=\textwidth]{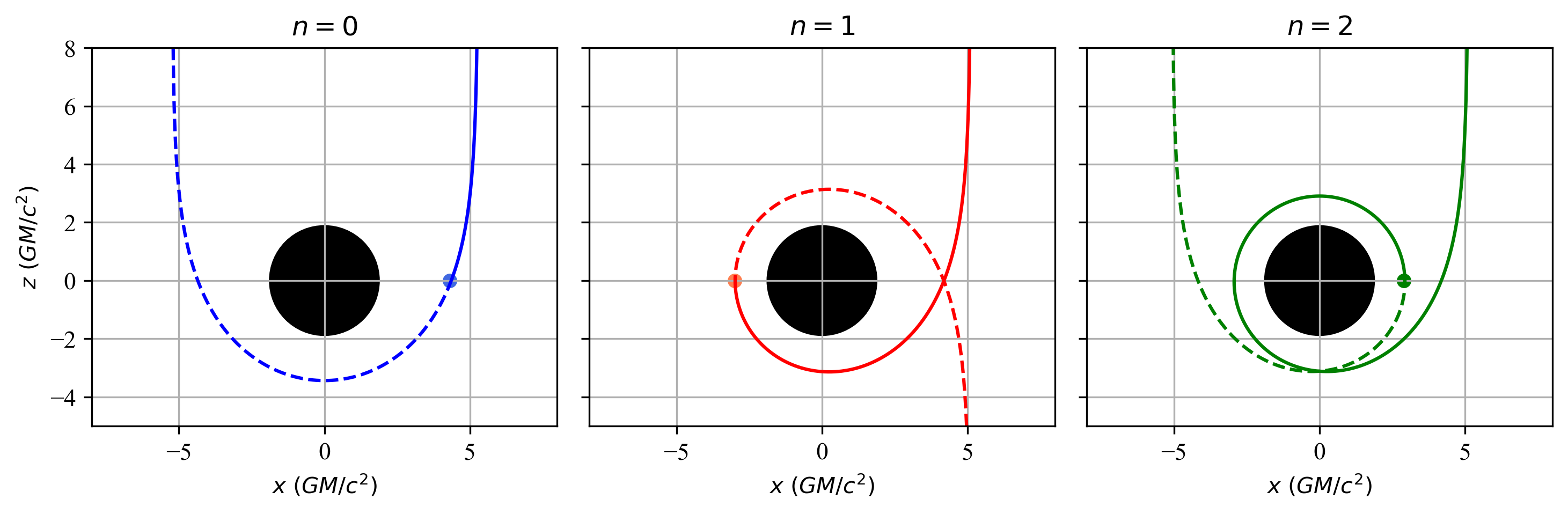}
    \caption{
    Photons orbit towards a polar observer, located at $z= \infty$, projected into the x-z plane in Kerr spacetime with
$a=0.5M$. The diagram shows the geometric picture of three distinct photon rings originating from the astrophysical disk around the black hole in the equtorial plane. On the right, corresponds to the direct image
$n =0$ ring. The $n=1$ ring which is a null ray that completes half an orbit around the black hole before escaping to infinity is shown in the middle. The $n=2$ ring is depicted by the green line on the left. The dark circle at the center represents the event horizon for $a=0.5M$ in Kerr spacetime. The dots show the location of emission on the equtorial plane and the dashed lines show the continuation of the trajectory to infinity. }
    \label{fig:geomttric pics of n=0,1,2}
\end{figure*}

\subsection{The Astrophysical Picture of Photon Rings}

\begin{figure}
    \centering
    \includegraphics[width=\columnwidth]{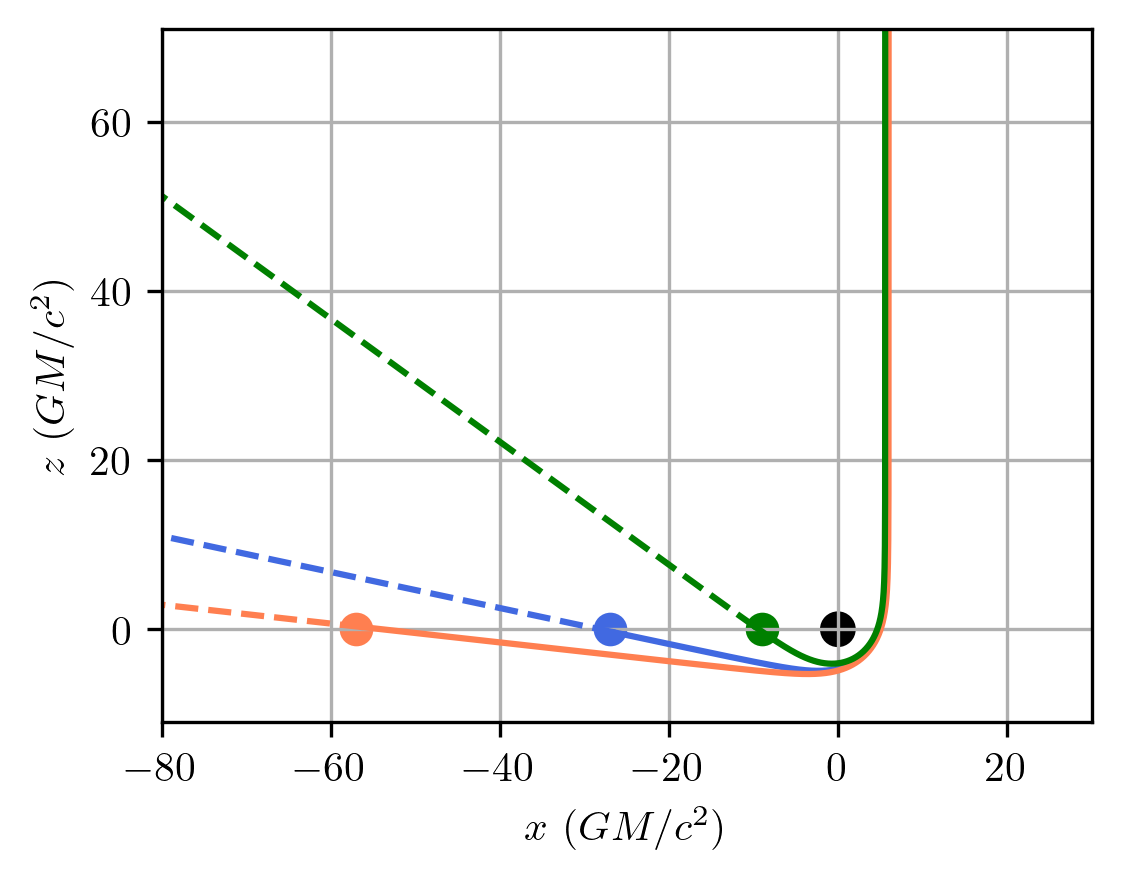}
    \caption{Null trajectories that travel towards the polar observer at different emission radii in the equtorial plane is shown here. The dashed lines represent the direct emission $n=0$ coming from $r_{em}=10M, 26M, 55 M$ and the solid lines represent the $n=1$ photon rings. The dark disk in the center demonstrates the horizon for Kerr spacetime with $ a=0.5 M$. This figure illustrates null trajectories originating from same location of emission can contribute to different subsequent rings depending on their initial conditions.}
    \label{fig:n=1 and 0 for diff rem}
\end{figure}

\begin{figure}
    \centering \includegraphics[width=\columnwidth]{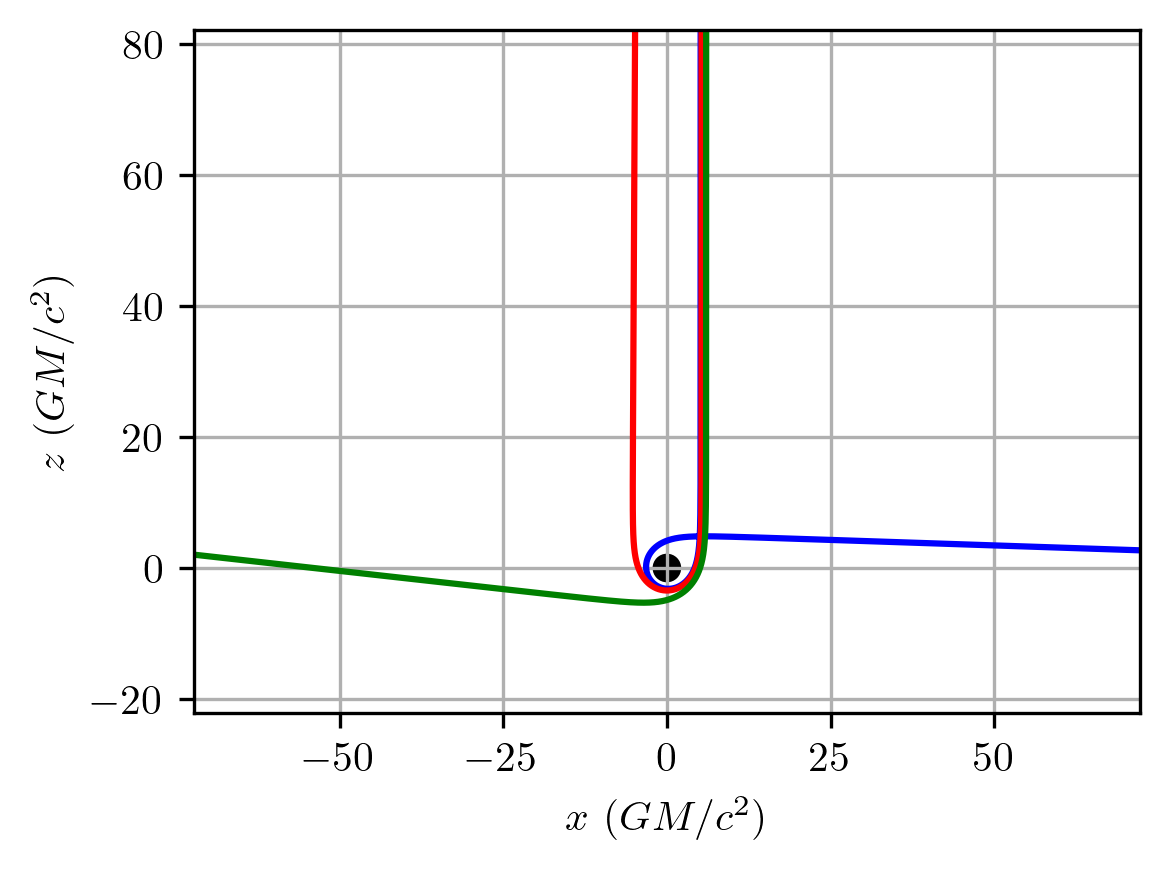}
    \caption{This visualization illustrates null trajectories approaching a polar observer from various emission radii within the equatorial plane. All these rays contribute to the $n=1$ photon ring. The blue trajectory completes three-quarters of an orbit around the black hole, the red trajectory represents a half orbit, and the green trajectory completes one-quarter of an orbit. These different trajectories offer insight into the geometric and dynamical characteristics of photon rings.}
    \label{fig:n=1 astrophysical picture}
\end{figure}

In realistic astrophysical scenarios, emission from different radii can contribute to multiple image orders. This emission usually is originated from the accretion disk surrounding the blackhole. As illustrated in \autoref{fig:n=1 and 0 for diff rem}, radiation from the same emission location may produce both the $n=0$ and $n=1$ images. This arises because photons from these regions undergo different total angular deflections along their trajectories. For example, in the case of $n=1$, the deflection angle is not exactly equal to $2\pi$, which is the geometric picture of $n=1$ orbit and is highlighted in \autoref{fig:geomttric pics of n=0,1,2}. Instead, all trajectories experiencing a deflection between $3\pi/2 + \epsilon$ and $5\pi/2 - \epsilon$, for arbitrarily small $\epsilon$, contribute to the $n=1$ image (see \autoref{fig:n=1 astrophysical picture}). This behavior can also be understood in light of Eq.~1 of \cite{Broderick2022}, which shows that the observed radius of the $n$-th order photon ring, $R^{obs}_n$, for a realistic emission region depends in a nontrivial way on the black hole mass, spin, and the emission radius.

\begin{equation}
\label{eq:theta_n}
    R^{obs}_n = (GM/c^2 D)  \vartheta_n(a,r_{em}c^2/GM),
\end{equation}
where M and a represent the mass and  spin of the black hole and D stands for the distance to the source. Therefore, the size of subsequent photon rings is a degenerate measurement of these parameters. After specifying the spin $a$ (and implicitly determining  $M$ through chosen units), the observed radii of photon rings constitute a one-dimensional continuum spanned by  $r_{em}$.

The event horizon sets a fundamental lower limit on $r_{em}$, but there is no upper bound on the emission radius. Therefore, there is no upper bound on the apparent radius of the $n=0$ rings in the images since photons from anywhere in the universe, even with slight deflection by the black hole's gravitational field, can contribute to the $n=0$ photon ring. 

However, this effect diminishes for higher-order rings. These rings experience greater lensing during their half orbits around the black hole before reaching the observer. The closer these rays travel to the critical radius $r_\gamma$ , the more they are lensed. For example, the $n=1$ ring is bound within 
$4.30< R^{obs}_{n=1} < 6.17$ , and for the 
$n=2$ ring, the limit changes to 
$4.77< R^{obs}_{n=2} < 5.22$. The limit gets more restrict drastically as the value of n increases \citep{Broderick2022}.


Therefore, unlike the geometric case, the observed radius of any subsequent image, $R^{\mathrm{obs}}_n$, is not a single value but spans a range that becomes narrower for higher-order photon rings, as mentioned earlier.  

In a realistic astrophysical environment, the Lyapunov exponent inferred from low-order rings---originating from null rays emitted at different radii $r_{\mathrm{em}}$---can be defined as
\begin{equation}
\label{eq:gamma obs n with R}
  \gamma^{\mathrm{obs},01\infty}_n = \ln \left( \frac{R^{\mathrm{obs}}_n - R_\infty}{R^{\mathrm{obs}}_{n+1} - R_\infty} \right),
\end{equation}
and
\begin{equation}
\label{eq:gamma obs n no R}
  \gamma^{\mathrm{obs},012}_n = \ln \left( \frac{R^{\mathrm{obs}}_n - R^{\mathrm{obs}}_{n+1}}{R^{\mathrm{obs}}_{n+2} - R^{\mathrm{obs}}_{n+1}} \right).
\end{equation}

These estimates deviate from the Lyapunov exponent defined in \autoref{eq:gamma} for two reasons:   

(i) the discrepancy between the idealized \textbf{geometric picture} and the \textbf{astrophysical picure}

(ii) the error due to usage of low-order rings to estimate the  Lyapunov exponent instead of higher order rings.

A critical question then arises: how much does the Lyapunov exponent estimated from lower-order rings deviate from the asymptotic value given by \autoref{eq:gamma}, which is derived in the limit of high-order photon rings?.This key issue will be addressed in the following section.

In this work for simplicity, we consider a hypothetical scenario where emission emanates from a single, azimuthally symmetric ring located in the equatorial plane at a radius $r_{em}$ . While the actual astrophysical environment may differ significantly from this idealized case, the radius of this emission ring can be substituted with the location of the emission maximum in models with extended emission \citep{Broderick2022}.

\section{ Convergance in the Photon Ring Radii}
\label{sec: Convergance in the Photon Ring Radii}

Having established the main sources of uncertainty in estimating the Lyapunov exponent, we now aim to quantify them. Specifically, we investigate (i) how much uncertainty arises from using the astrophysical values of the image radii rather than the geometric ones, and (ii) how the use of low-order rings affects the accuracy of our estimates.  

\subsection{Geometric vs. Astrophysical Picture}

In this subsection, we demonstrate how $R^{\mathrm{obs}}$ approaches $R^{\mathrm{geo}}$ as $n$ increases. Importantly, this convergence occurs at precisely a rate which is sufficient to improve the estimates of the Lyapunov exponent.  

\begin{figure*}[!t]
    \centering
    \includegraphics[width=\textwidth]{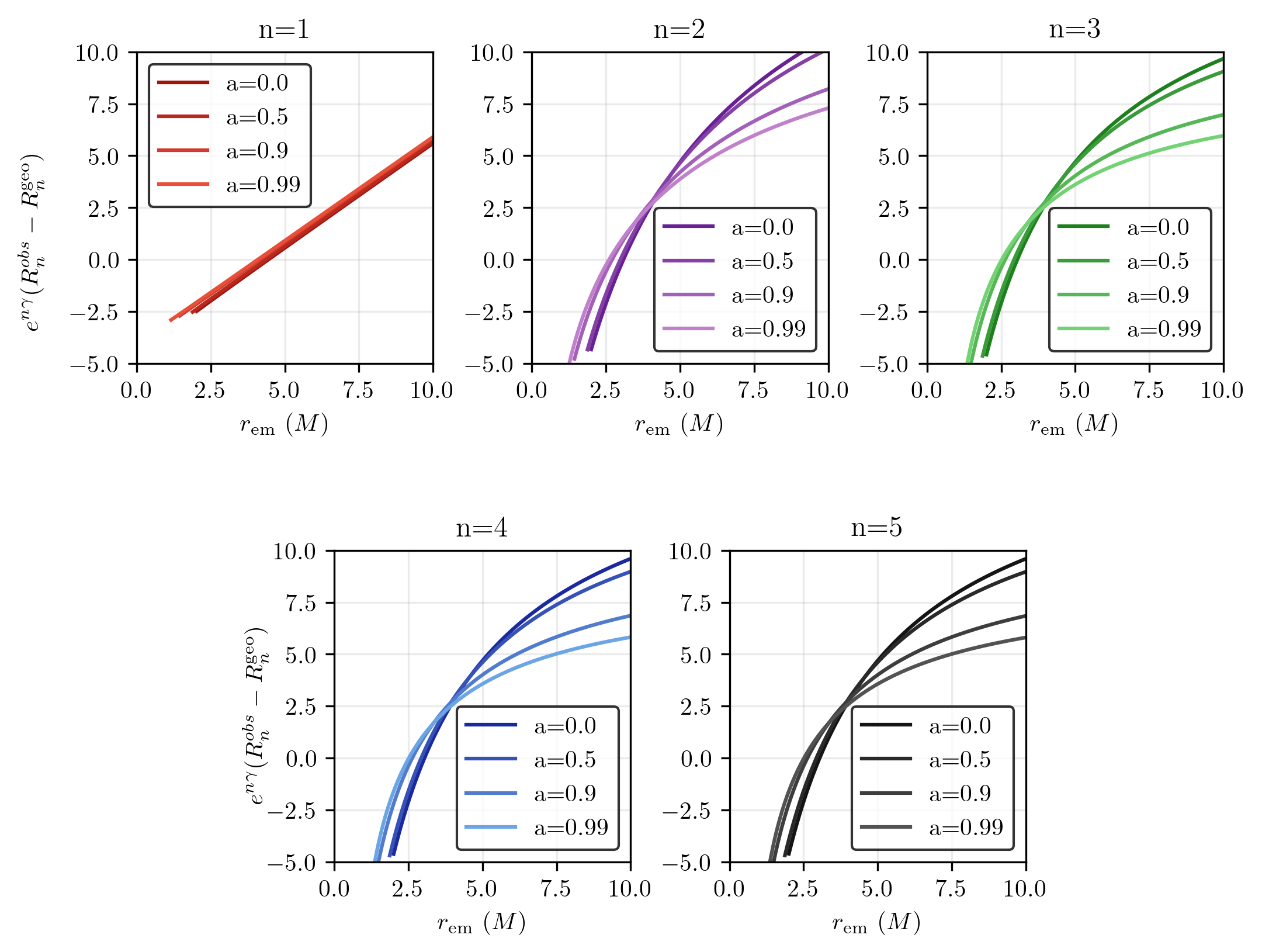}
    \caption{
    The y-axis represents the difference between the radius of the $n$th photon ring originating from $r_{em}$ and the corresponding geometric values of the apparent radius for various spin parameters and ring orders. The leftmost plot shows this difference for the secondary image with $n=1$, and as you move rightward, the value of $n$ increases, with the rightmost plot representing $n=5$}
    \label{fig:R-Rge_diff_a_and_n}
\end{figure*}

As illustrated in \autoref{fig:R-Rge_diff_a_and_n} for different spin values, $R^{\mathrm{obs}}$ rapidly converges toward $R^{\mathrm{geo}}$. The difference between the geometrical and observed value of photon ring for $n=1$ is highly influenced by the astrophysics (see top left of the \autoref{fig:R-Rge_diff_a_and_n}). Moreover, as $n$ increases this effect gets less significant and the curves tend to reflect on the purely gravitational effects. Therefore, for $n=2$, the curve is significantly different than $n=1$ and much closer to $n=3$'s diagram.
For $n \geq 3$, the functional form of the curves is nearly identical to each other, differing only by an exponential offset. This exponential convergence ensures that the inferred Lyapunov exponent from $R^{\mathrm{obs}}$ improves systematically with ring order. Thus, by applying \autoref{eq:gamma012} and \autoref{eq:gamma with R}, one can deduce that $\gamma^{\mathrm{obs}}$ converges to $\gamma^{\mathrm{geo}}$.

\subsection{Low-Order Ring Estimates}

\begin{figure}
    \centering   
    \includegraphics[width=\columnwidth]{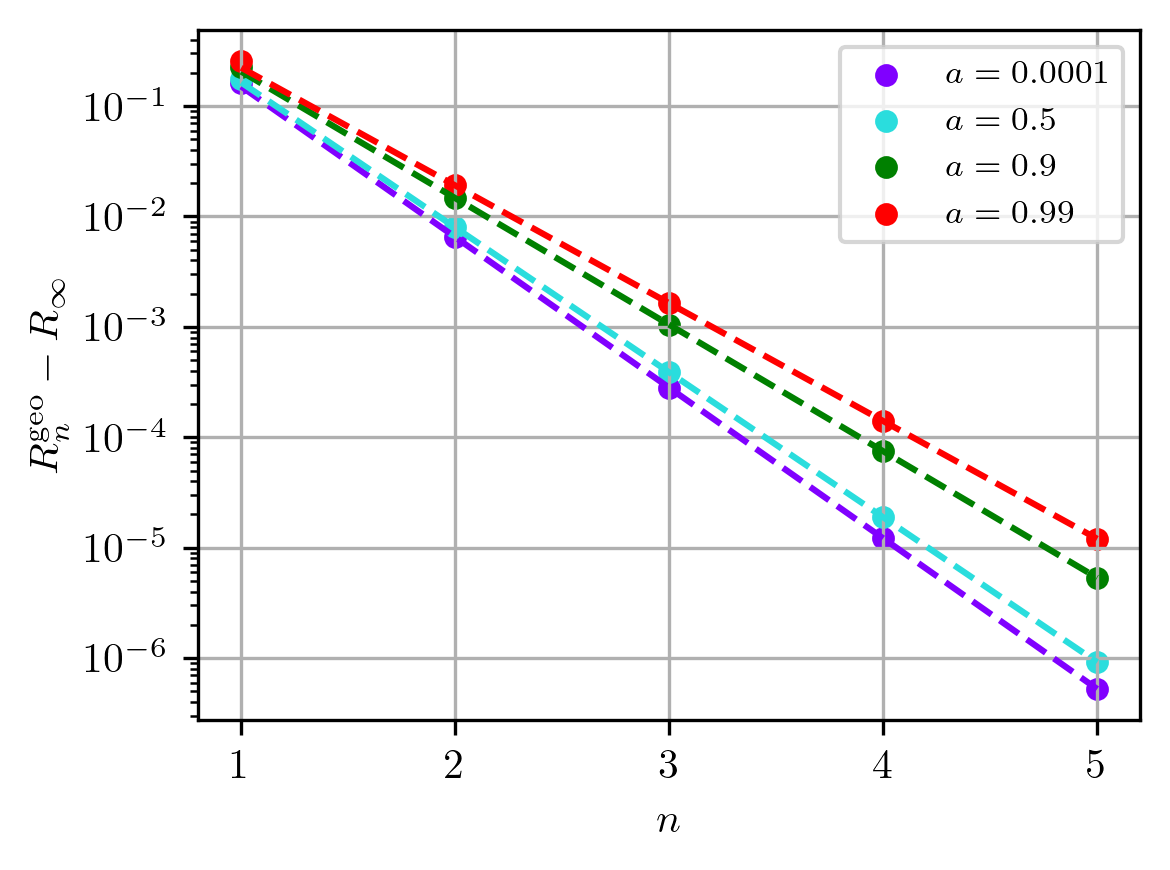}
    \caption{The y-axis represents in this plot illustrates the difference between the geometric radius of each photon ring and the shadow size, presented on a logarithmic scale. The x-axis represents the order of the photon ring, causing the points to be scattered along the plane. The lines depict $\Delta R e^{n \gamma}$ for various spin values, and because the y-axis is logarithmic, these appear as straight lines. }
    \label{fig: Rgeo - shadow}
\end{figure}

Although we have shown that estimates of the Lyapunov exponent based on observed photon ring radii converge to their geometric counterparts, we must also determine whether the geometric Lyapunov exponent itself converges to the asymptotic expression in \autoref{eq:gamma} as $n$ increases.  

This behavior is explored in \autoref{fig: Rgeo - shadow}. As shown, the geometric ring radius $R^{\mathrm{geo}}$ approaches the shadow radius exponentially with increasing $n$, as indicated by the dashed lines. Additionally, this behavior was expected due to \autoref{eq:Rn and shadow}. Since $R^{\mathrm{geo}}$ takes a unique value for each $n$, it is represented by solid dots in the figure, which align precisely along the $\Delta R e^{-n \gamma}$ curves. This alignment confirms the exponential convergence of $R^{\mathrm{geo}}$ to the shadow radius, exactly as required for consistency with \autoref{eq:gamma}.  

Combining this result with the previous subsection, we conclude that despite the two main sources of error—(i) the use of low-order rings and (ii) the discrepancy between the geometric and astrophysical pictures—the observed estimates $\gamma^{\mathrm{obs}}$ converge to the asymptotic Lyapunov exponent defined in \autoref{eq:gamma}.

However, the practical question of how rapid this convergence is remains unanswered. We address this issue in the following section. 

\section{Measuring the Lyapunov Exponent}
\label{sec:Measuring gamma}

In the previous sections, we discussed the challenges in accurately estimating the Lyapunov exponent due to two main factors: first, the discrepancy between the geometric and astrophysical perspectives, and second, how the Lyapunov measurement using the geometric model still differs from the theoretical value using the low order photon rings. We demonstrated that these issues become less significant when using higher-order photon rings rather than lower-order ones. However, one might still question how quantitatively inaccurate the estimation of $\gamma$ is when relying on low-order photon rings. In this section, we will explore this issue. 

\begin{figure}
    \centering
    \includegraphics[width=\columnwidth]{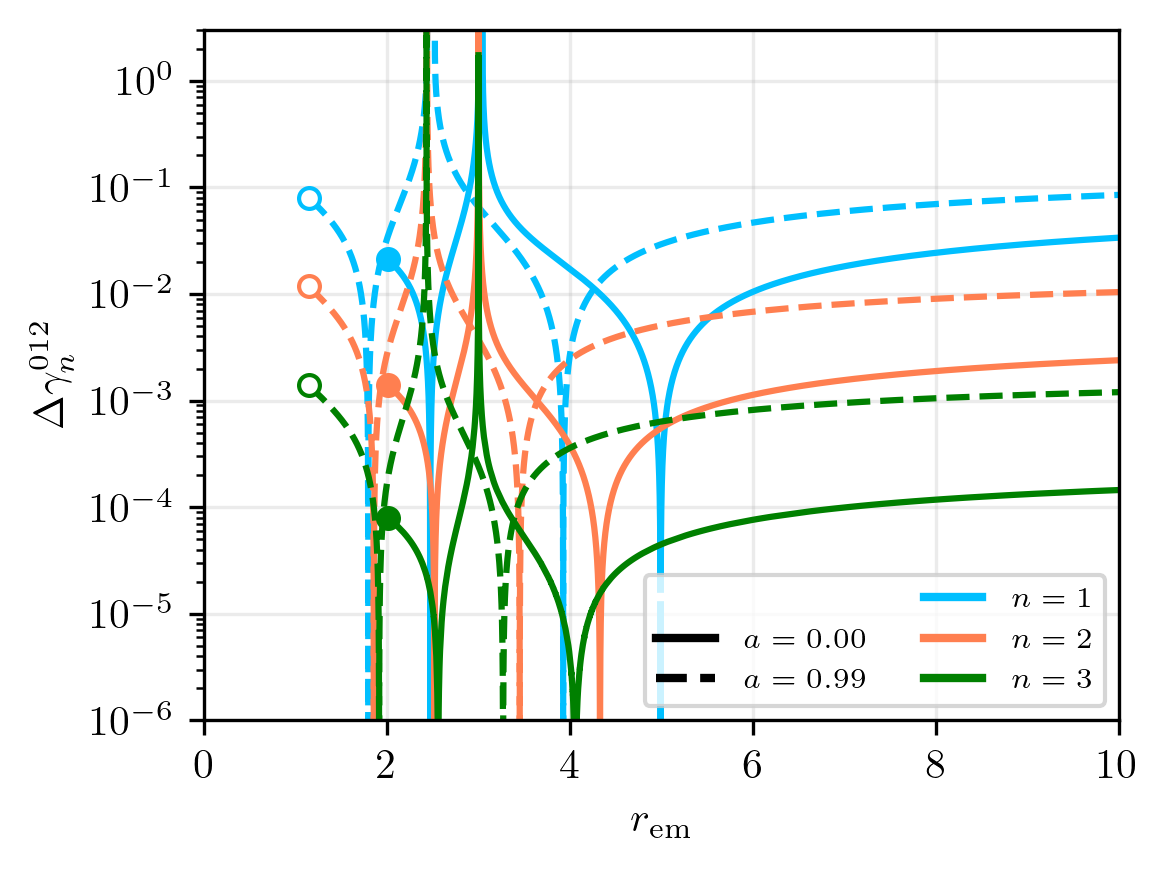}
    \caption{Fractional errors in estimating the Lyapunov exponent by using the radii of low-order photon rings using \autoref{eq: Delta gamma obs} for  \textbf{maximally spinning} and \textbf{non- spinning} black hole spacetimes. This estimation of Lyapunov exponent is without using the shadow size. The different  orders of photon rings are shown in different color and the difference between the maximally spinning and non spinning black hole spacetime is shown by different shades.}
    \label{fig:Delta gamma 012}
\end{figure}

\begin{figure}
    \centering
    \includegraphics[width=\columnwidth]{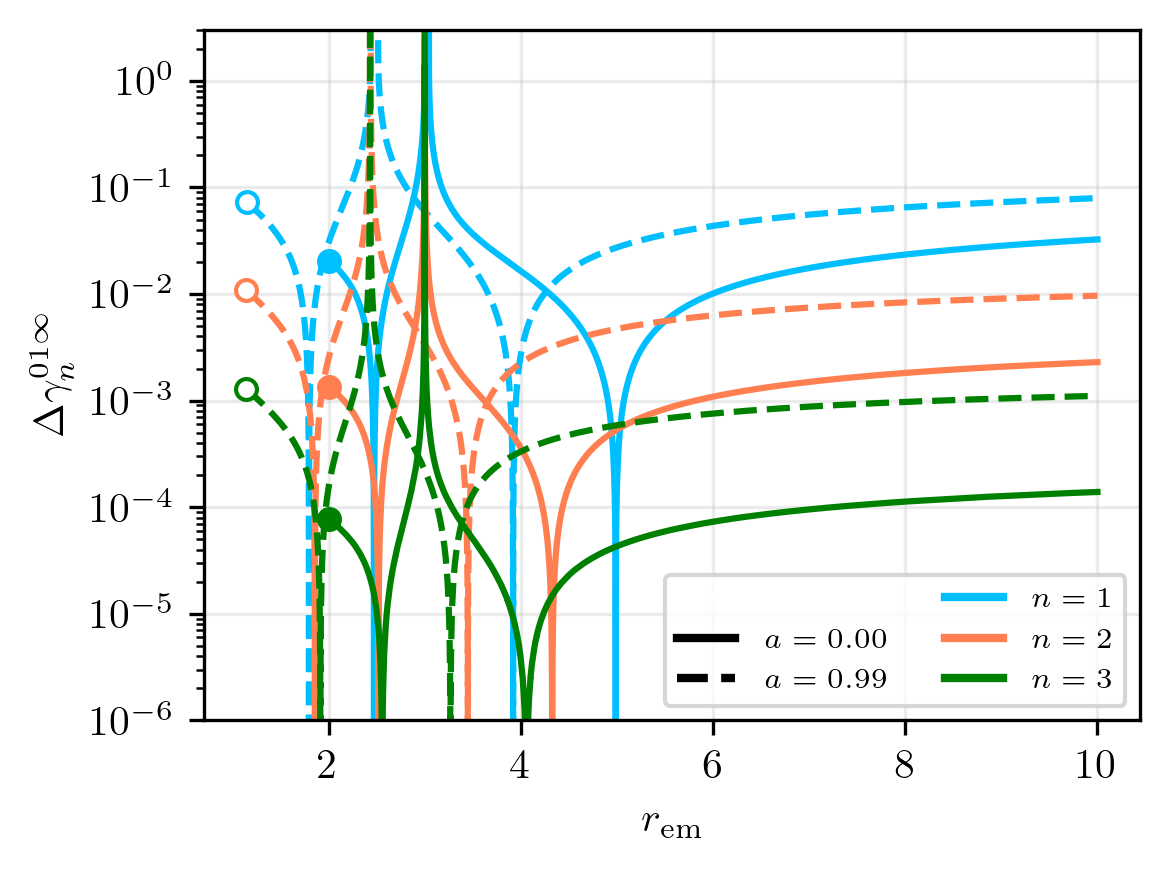}
    \caption{\textbf{change the labels, titles} Fractional errors in estimating the Lyapunov exponent by using the radii of low-order photon rings and the shadow size using \autoref{eq: Delta gamma obs} for  \textbf{maximally spinning} and \textbf{non- spinning} black hole spacetimes. This estimation of Lyapunov exponent is without using the shadow size. The different  orders of photon rings are shown in different color and the difference between the maximally spinning and non spinning black hole spacetime is shown by different shades.}
    \label{fig:Delta gamma 01 inf}
\end{figure}

In order to quantify these errors in the estimate of the Lyapunov exponent, we define the error associated with \textbf{geometric} picture (\autoref{eq:gamma geo n with R} and \autoref{eq:gamma geo n no R}) to be, 
\begin{equation}
    \Delta \gamma^{geo,012}_{n}  = \frac{\gamma^{geo,012}_{n}- \gamma}{\gamma} ~~~\text{and}~~~ \Delta \gamma^{geo,01\infty}_{n} = \frac{ \gamma^{geo,01\infty}_{n}- \gamma}{\gamma}.
    \label{eq: Delta gamma geo obs}
\end{equation}
which, here $\gamma$ is the Lyapunov exponent coming from \autoref{eq:gamma}, which is for the case of very high order rings. 
Similarly, the discrepancy between the the \textbf{astrophysical picture} (\autoref{eq:gamma obs n with R} and \autoref{eq:gamma obs n no R}) and $\gamma$, which is the ultimate uncertainty of interest, is defined by, 

\begin{equation}
    \Delta \gamma^{012}_{n}  = \frac{\gamma^{obs,012}_{n}- \gamma}{\gamma} ~~~\text{and}~~~ \Delta \gamma^{01\infty}_{n} = \frac{ \gamma^{obs,01\infty}_{n}- \gamma}{\gamma}.
    \label{eq: Delta gamma obs}
\end{equation}
These errors are shown in \autoref{fig:Delta gamma 012} and \autoref{fig:Delta gamma 01 inf}. 

These plots clearly illustrate that as higher photon orders are incorporated, the characteristic fractional error with both methods ($\Delta \gamma^{012}_{n}$ and $\Delta \gamma^{01\infty}_{n}$) typically decreases for both maximally spinning and non-spinning cases. However, there are a handful of values of $r_{\rm em}$ at which this trend is narrowly violated.  The locations associated with the anomalous behavior depend on black spin and the order of the photon rings employed.  When $R_{n}(r_{\rm em})$ becomes equal to the the shadow size (\autoref{fig:Delta gamma 01 inf} or $R_{n+1}(r_{\rm em})$ (\autoref{fig:Delta gamma 012}), the estimates of $\gamma$ diverge, resulting in catastrophic loss of accuracy over a narrow range of $r_{\rm em}$.  In contrast, the estimators of $\gamma$ become exact at two $r_{\rm em}$.  We don't expect either to present a significant impediment to measuring $\gamma$ in practice for sources with variable emission regions, and therefore sampling multiple $r_{\rm em}$.  
Thus, we concluded that the error typically improves exponentially as $n$ increases, being noticeable lower in the non-spinning spacetime.

\begin{figure}
    \centering
    \includegraphics[width=\columnwidth]{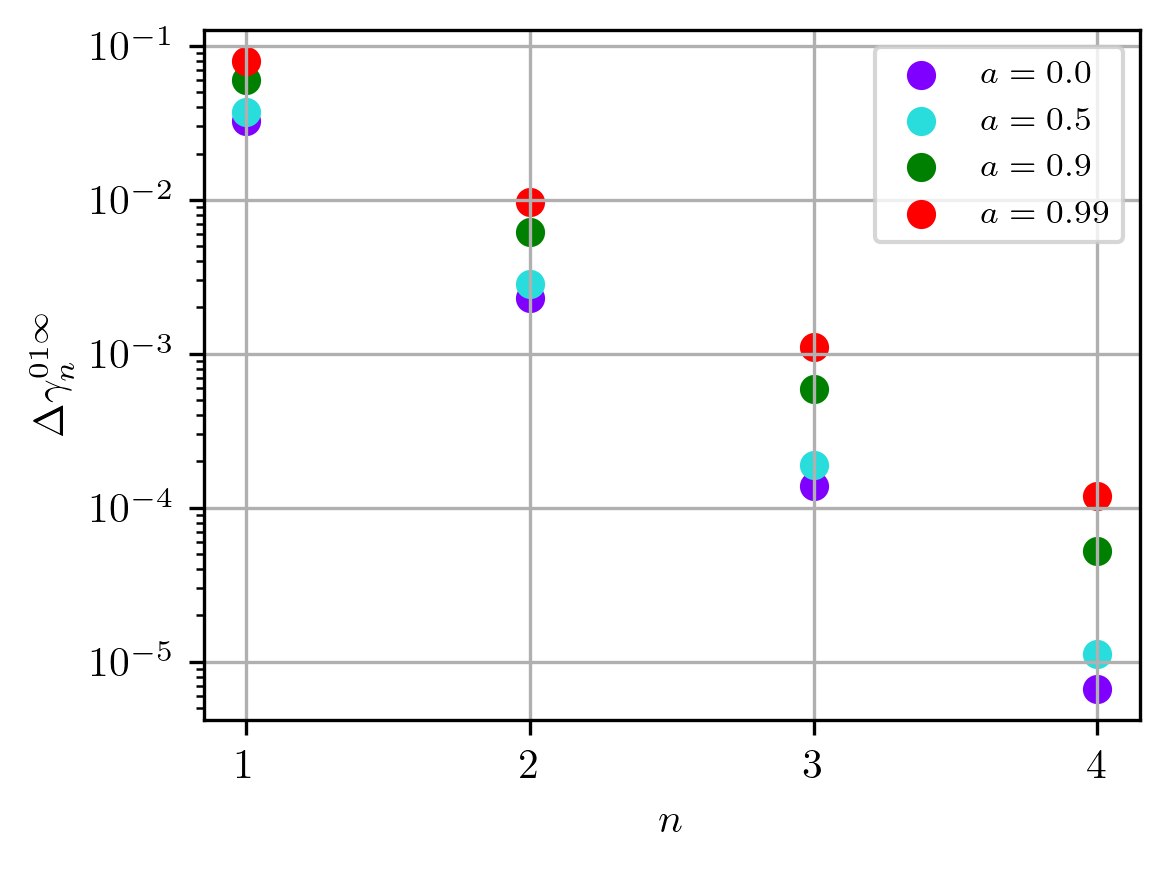}
    \caption{Fractional errors in estimating the Lyapunov exponent by using the radii of low-order photon rings and the shadow size for different values of black hole spin and photon ring orders. This Fractional errors are measured at $r_{em} = 10M$ as a reference point. }
    \label{fig:Delta gamma 01 inf at 10M}
\end{figure}

\begin{figure}
    \centering
    \includegraphics[width=\columnwidth]{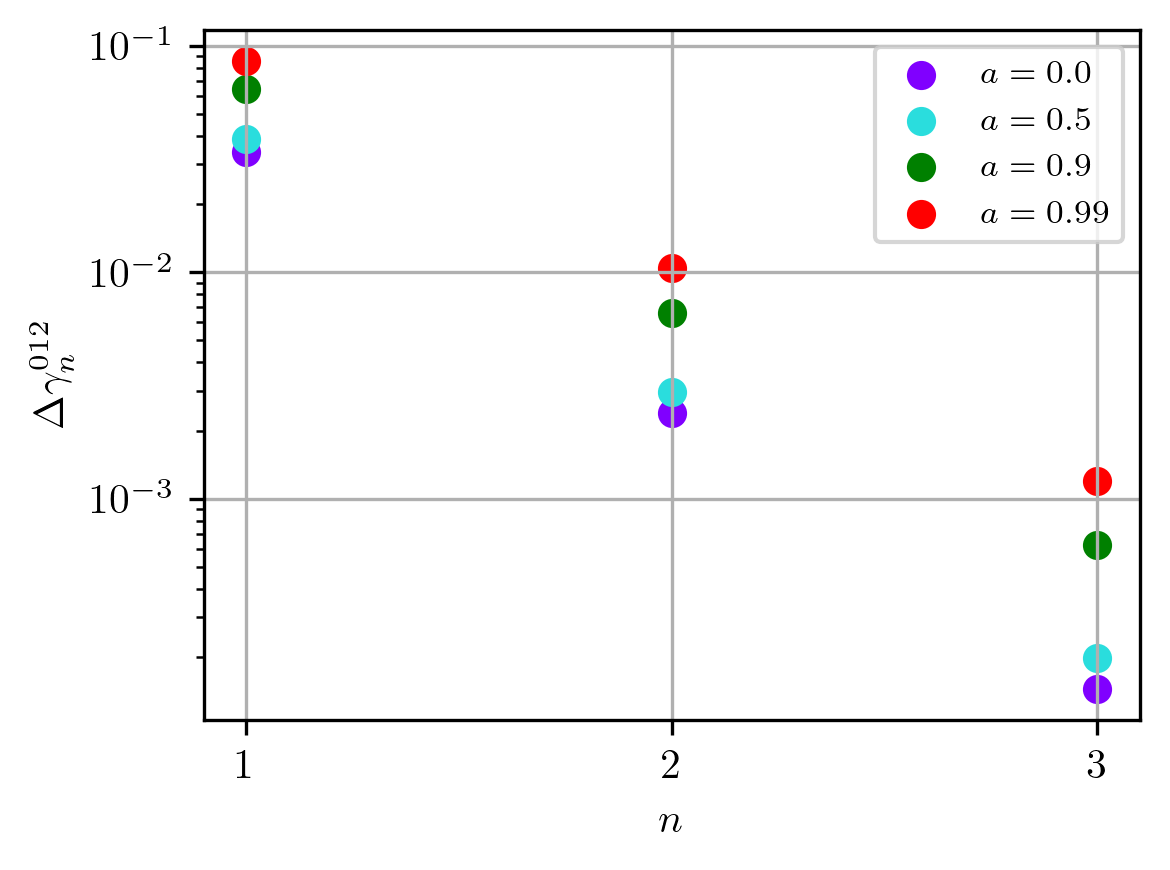}
    \caption{Fractional errors in estimating the Lyapunov exponent by using the radii of low-order photon rings for different values of black hole spin and photon ring orders. This Fractional errors are measured at $r_{em} = 10M$ as a reference point. }
    \label{fig:Delta gamma 012 at 10M}
\end{figure}

To quantify the fractional error for different photon orbit orders and spin values, we choose $\Delta \gamma^{01\infty}_n$ at $r_{\rm em} = 10M$ as a convenient reference point. This value lies in the middle of the fractional error intervals, as shown in \autoref{fig:Delta gamma 012} and \autoref{fig:Delta gamma 01 inf}. it is neither exceptionally accurate nor highly inaccurate compared to other emission radii. 
Because the fractional accuracy is insensitive to $r_{\rm em}$ outside of $\sim 5M$, the rate at which the estimates of $\gamma$ approach the true value are similarly insensitive to the reference $r_{\rm em}$ chosen.

This choice of a fiducial $r_{\rm em}$ is further motivated by the fact that in GRMHD simulations the emission is dominated by material within $10M$ \citep{M87_PaperV}.
\autoref{fig:Delta gamma 01 inf at 10M}  and \autoref{fig:Delta gamma 012 at 10M} present the fractional error in estimating the Lyapunov exponent for different spin values and photon ring orders. They show that estimates of $\gamma$ become exponentially better with the minimum order of photon ring used. The rate at which these estimates converge scales roughly as $10^{-n}$ for all spin values only changing mildly with the spin value. Nevertheless, by $n=2$ sub-percent accuracies are generally obtained.

For the sake of brevity, we do not show the plots for 
$\Delta \gamma^{\rm geo,012}_n$ and $\Delta \gamma^{\rm geo,01\infty}_n$. 
Nevertheless, we have examined that these quantities exhibit patterns 
similar to those in \autoref{fig:Delta gamma 012} and \autoref{fig:Delta gamma 01 inf}, 
as can be inferred from \autoref{fig:R-Rge_diff_a_and_n}. 
As described previously, in these plots, as $n$ increases, the observed image radii 
and the corresponding geometric image radii approach each other exponentially. 
This behavior precisely matches the rate required to correctly estimate the Lyapunov exponent 
(see \autoref{eq: Delta gamma geo obs}).

\section{Conclusions}
\label{sec:conclusions}
We explore potential sources of error in the measurement of the Lyapunov exponent for radial photon trajectories near the photon shell.  To ensure a well formulated null test, we do this solely in the Kerr spacetime.  Because M87* is the first target for which these estimates are likely to be practically possible, we further restrict our attention to polar observers, roughly consistent with the viewing angle of that source.

We define two estimators of $\gamma$ based upon a fully geometric picture of the photon rings, previously exploited in \citet{Broderick2023} and \citet{Salehi2023}. Two main sources of potential inaccuracy in estimates of $\gamma$ were quantitatively assessed:
\begin{enumerate}
    \item That arising from relying on low-order photon rings, for which the asymptotic behaviors may be violated at order unity,
    \item That due to the distinction between the geometric picture of higher order images and the more relevant astrophysical one in which the emission can arise from an extended region.
\end{enumerate}
The first is motivated by practical aspect of observations with EHT, ngEHT, and future instruments, all of which will be likely to only access the lowest order photon rings in the near future.  The second is a key systematic uncertainty associated with the practical uncertainty in the source and location of the emission observed.

We find that the geometric picture of photon rings, defined by half orbits about the black hole, and a more astrophysically relevant picture based on observed image order, are strongly correlated.  In the limit of high photon ring or image order, the two notions converge to identical photon ring radii.

Of particular importance for measuring $\gamma$, the rate of convergence of these two notions of the photon rings is faster than magnification due the lensing compresses subsequent images against the edge of the black hole shadow.  In this sense, the geometric picture of photon rings employed by \citet{SphericalShadows23} and \citet{Salehi2023} becomes exponentially more accurate with photon ring order.  For the astrophysical picture of photon rings, this convergence results in the development of an asymptotic functional relationship between emission radius and image location by photon ring order $n=3$.  Higher order images may be mapped on to each other by simple exponential stretch/compression relative to the geometric photon ring.

Because the error in locations of the photon rings decreases faster than gravitational lensing compresses the images, both of our estimators for $\gamma$ become exponentially more accurate as well.  This is true independent of the estimator chosen: $\gamma_n^{01\infty}$ which makes use of the shadow size and two photon ring orders, or $\gamma_n^{012}$ which uses three photon ring orders but not the shadow.  This improvement occurs independent of the spin of the black hole, scaling faster than $10^{-n}$ for all spins and estimators, where $n$ is the lowest-order photon ring included.

The estimates of $\gamma$ are sensitive to the location of the emission region within the equatorial plane.  For each set of photon ring orders, there are specific emission radii at which the accuracy becomes uncharacteristically good/poor.  This behavior is confined to narrow regions in $r_{\rm em}$, and thus, for a variable source is unlikely to seriously impact efforts to measure $\gamma$.  

Even low-order photon ring measurements can produce highly accurate estimate of $\gamma$.  For example, $\gamma$ may be estimated to 1\% if the $n=2, 3$ and either the $n=4$ photon rings or the shadow size are accurately measured.  Even measuring only the $n=1, 2$ and either $n=3$ photon rings or shadow size achieves an accuracy of 10\%.  Therefore, these results should motivate the design of forthcoming black hole imaging efforts to capture, even if indirectly, the $n=2$ photon ring.

Three obvious limitations of our study that could be relaxed in future studies include the assumption of the polar observer, the equatorial nature of the emission region, and the restriction to the Kerr spacetime.  While each is well justified by existing observations of M87*, clear dividends may be found by extending any.  Of these, the viewing angle is the most natural, providing access to asymmetry imposed by black hole spin.  The small inferred inclination of M87*'s spin means that an expansion about the polar observer may be particularly profitable, similar to that employed by \citet{Salehi2025}.  

Emission from above and below the equatorial plane is often invoked in simulations of horizon-resolving images of M87*.  It is, nevertheless, usually symmetric about the equatorial plane, again suggesting a perturbative approach may be helpful, providing information about the latitude dependence of $\gamma$.  

Our restriction to Kerr spacetimes is well motivated by our desire to construct a null test.  However, this does prevent the characterization of how large a deviation might be theoretically expected, or nonlinear phenomena that may appear close to the photon orbit.  Existing libraries of alternative spacetimes and general frameworks in which they are implemented exist \citep[e.g.,][]{Salehi2023}, and point the way for repeating the analysis we present here.

We did not evaluate the impact of observational uncertainties.  The practical application of our estimators should be accompanied by validation experiments that incorporate instrumental properties and observational strategy.  In practice, we expect the precision with which $\gamma$ may be measured will be limited by the accuracy with which the highest order photon ring radii may be determined.  Importantly, we conclude that systematic uncertainties do not preclude measuring the shape of the Kerr spacetime near the photon orbit.

\begin{acknowledgments}
\section{acknowledgments}
The Perimeter Institute for Theoretical Physics partially supported this work. Funding for research at the institute is provided by the Department of Innovation, Science and Economic Development Canada, and the Ministry of Economic Development, Job Creation and Trade of Ontario, both of which are branches of the Government of Canada.  Additionally, A.E.B. receives further financial support for this research through a Discovery Grant from the Natural Sciences and Engineering Research Council of Canada.
\end{acknowledgments}

\vspace{5mm}

\clearpage

\newpage
\bibliographystyle{aasjournal}
\bibliography{references.bib}

\end{document}